\documentclass[journal]{IEEEtran}
\usepackage{amsmath, amsthm, amsfonts}
\usepackage[english]{babel}
\usepackage{graphicx}
\usepackage{tabularx}
\usepackage{multirow}
\usepackage{enumerate}
\usepackage{setspace}
\usepackage[ruled,vlined]{algorithm2e}
\usepackage{caption}

\begin{document}

\title{VLSI Friendly Framework for Scalable Video Coding based on Compressed Sensing}

\author{{B.K.N.Srinivasarao$^1$, Vinay Chakravarthi Gogineni $^2$, Subrahmanyam Mula $^3$ and Indrajit Chakrabarti $^4$}\\
Department of Electronics and Electrical Communication Engineering\\
Indian Institute of Technology, Kharagpur, INDIA\\
E.Mail : $^1\;$srinu.bkn@iitkgp.ac.in,
$^2\;$vinaychakravarthi@ece.iitkgp.ernet.in,
$^3\;$subrahmanyam.mula@gmail.com,
$^4\;$indrajit@ece.iitkgp.ernet.in }
\maketitle
\thispagestyle{empty}
\begin{abstract}

 This paper presents a new VLSI friendly framework for scalable video coding  based on Compressed Sensing (CS). It achieves scalability through 3-Dimensional Discrete Wavelet Transform (3-D DWT) and better compression ratio by exploiting the inherent sparsity of the high frequency wavelet sub-bands through CS. By using 3-D DWT and a proposed adaptive measurement scheme called AMS at the encoder, one can succeed in improving the compression ratio and reducing the complexity of the decoder. The proposed video codec uses only 7$\%$ of the total number of multipliers needed in a conventional CS-based video coding system. A codebook of Bernoulli matrices with different sizes corresponding to the predefined sparsity levels is maintained at both the encoder and the decoder. Based on the calculated $\ell_0 $-norm of the input vector, one of the sixteen possible Bernoulli matrices will be selected for taking the CS measurements and its index will be transmitted along with the measurements. Based on this index, the corresponding Bernoulli matrix has been used in CS reconstruction algorithm to get back the high frequency wavelet sub-bands at the decoder. At the decoder, a new Enhanced Approximate Message Passing (EAMP) algorithm has been proposed to reconstruct the wavelet coefficients, and apply the inverse wavelet transform for restoring back the video frames. Simulation results have established the superiority of the proposed framework over the existing schemes  and have increased its suitability for VLSI implementation. Moreover, the coded video is found to be scalable with increase in number of levels of wavelet decomposition.

\end{abstract}


\section{Introduction}

\par Current video coding standards (e.g. H.264 and HEVC)\cite{H.264}\cite{HEVC} depend on high-complexity encoder to provide good compression. They are more suitable for the broadcasting applications, where a highly complex encoder would support thousands of moderately complex decoders. However, conventional video coding schemes are not suitable for applications like mobile phones and camcorders which require low complexity encoders. Low complexity coding algorithms are desirable for reducing the demand on on-board resources and increasing the down-link transmission rate of spacecraft in space science applications \cite{space}. Therefore to reduce the power consumption, a low-complexity encoder with good coding efficiency is highly desirable. 
  
\par For mobile streaming and broadcasting application, different users may have varied processing power, network bandwidth and demand for different versions of the same content. This fact is addressed in Scalable Video Coding (SVC) \cite{CS_1}\cite{CS_2}, which enables decompression at various frame rates and resolutions as specified by the user terminal, from the video stream which was compressed with high  resolution at the encoder.  SVC offers scalability in three important dimensions, viz. spatial, temporal and quality and facilitates a flexible solution for transmission over diversified networks.

\par Based on the ideas of compressed sensing (CS) \cite{CS_11}-\cite{CS_13}, several new video codecs \cite{CS_10}-\cite{CS_plonka} have been proposed in the last few years and are suitable for application of wireless networks. Wakin \textit{et al.} \cite{CS_10} have introduced the compressive imaging and video encoding through single pixel camera. From his research results, Wakin has established that 3-D wavelet transform is a better choice for video compared to 2-D (two-dimensional) wavelet transform. Most of the recently proposed video codecs \cite{CS_10}-\cite{CS_plonka}, which are assumed to be of uniform sparsity, are available for all the video frames and a fixed number of measurements are transmitted to the decoder for all the frames. Depending on the content of the video frame, sparsity may change. A fixed number of measurements may cause an increase in bit-rate (decrease in compression ratio). 

\par Although some of the existing CS based video coding schemes provide good reconstruction quality \cite{31}-\cite{CS_plonka}, they assume sparsity in any of the transform domain and multiply the video cubes with randomly generated sensing matrix at the encoder before transmission. Whereas, the decoder involves complex operations like forward and inverse transformation (preferably DCT or DWT) and multiplication of sensing matrix (both original matrix and inverse matrix) with transform coefficients, which makes the decoder highly complex, while on the other hand, encoder is much less complex. Moreover, the hardware implementation of such high complexity decoders is very involved. Therefore, it is desired to have video coding system with low complexity, good coding efficiency, error resilience, scalability, and good quality. Moreover, to support  real-time processing, it should be amenable to hardware implementation.

\par
This paper proposes a novel SVC framework that achieves scalability through 3-D wavelets and better compression ratio by exploiting the inherent sparsity of the high frequency wavelet sub-bands through CS. At the encoder, 3-D DWT has been applied on group of frames (GOF), and multiple columns of each sub-band are converted to single column input vector and the $\ell_0 $-norm $(K)$ is subsequently calculated. Unlike the conventional CS, which takes uniform number of measurements for all the input vectors, 3-D DWT in the proposed framework enables the adaptive measurement scheme (AMS) by calculating the exact sparsity ($\ell_0 $-norm). This invariably leads to an increase in the compression ratio compared to the existing CS based video coding schemes. A codebook of sensing matrices (Bernoulli matrices) with different sizes corresponding to the predefined sparsity levels is maintained at both the encoder and the decoder. Based on the calculated $\ell_0 $-norm of the input vector, one of the Bernoulli matrices will be selected for taking the CS measurements and its index will be transmitted along with the measurements. Based on this index, the corresponding Bernoulli matrix (sensing matrix) has been used in CS reconstruction algorithm to get back the high frequency wavelet sub-bands at the decoder. To improve the reconstruction quality over Approximate Message Passing (AMP) algorithm \cite{dono_1}\cite{dono_2}, an enhanced version of AMP (viz. EAMP) has been proposed in this work. Use of 3-D DWT at encoder significantly reduces the complexity of the decoder in comparison with conventional CS based decoder and is shown in Table \ref{Comp_compare}. It shows that for 25$ \% $ measurements, nearly 94$ \% $ of multipliers are reduced involving however an increase of 6$ \% $ adders in the proposed framework compared to the conventional CS based video coding systems and hence it is more suitable for VLSI implementation which will be explained in the later section. 
\begin{table*}[]
\centering
\caption{Complexity Comparison of the proposed system with conventional system}
\label{Comp_compare}
\begin{tabular}{|l|l|l|}
\hline
                            & Conventional CS based Video codec                                        & Proposed CS based Video codec                                                                             \\ \hline
Encoder                     & $y=\Phi x$                                                               & $y=\Phi (\Psi^T x)$   \\ \hline
\multirow{2}{*}{Complexity} & \multirow{2}{*}{No. of Adders: M$ \times $(N-1)  }                                & No. of Multipliers: 6$ \times $N$ \times $N  \\ 
\cline{3-3} 
                            &                                                                          &No. of Adders: 6(N$ \times $(N-1))+(M$ \times $N-1) \\ \hline
                            
Decoder                     & $\hat s(n + 1) = \hat s(n) + {(\Phi \Psi )^T}(y - (\Phi \Psi )\hat s(n)$ & $\hat s(n + 1) = \hat s(n) + {\Phi^T}(y - \Phi\hat s(n)$             \\ \hline

\multirow{2}{*}{Complexity} & No. of Multipliers: (2$ \times $M$ \times $N)$ \times $Itr  & \multicolumn{1}{c|}{\multirow{2}{*}{No.  of Adders:     ((M$ \times $N-1)+ (N$ \times $M-1))$ \times $Itr}} \\ \cline{2-2}
                            & No. of Adders : ((M$ \times $N-1)+ (N$ \times $M-1))$ \times $Itr & \multicolumn{1}{c|}{}                                                                \\ \hline
      Total Complexity with 25$ \% $ & No. of Multipliers= 0.5$N^{2} \times $Itr  & No.  of Multipliers $\approx $  6$N^{2}$ \\ \cline{2-3}
                    Measurements (M=0.25N)        & No.  of Adders $\approx $  0.5$N^{2} \times $Itr  & No. of Adders $\approx $ (0.5$N^{2} \times $Itr)+ 6.25$N^{2}$ \\ \hline      
Total Complexity with & No. of Multipliers= 100$N^{2}$ & No.  of Multipliers $\approx $  6$N^{2}$ \\ \cline{2-3}
                   M=0.25N and Itr = 200        & No.  of Adders $\approx $  100$N^{2}$  & No. of Adders $\approx $106.25$N^{2}$ \\ \hline                           
                                              
                            \multicolumn{3}{l}{N = No.of elements in a video cube or GOF, Itr = No.of iterations, which is around 200, size of $\Phi$ is M$ \times $N} \\
                            \multicolumn{3}{l}{$\Psi^T$ is a lifting based wavelet transform, size of $\Psi$ is N$ \times $N }                                                                                                                                                                                                                                                                                                                                                       \\
\end{tabular}
\end{table*}
\par The organization of the remaining paper is as follows. Review of compressed sensing based video coding is provided in Section II. Detailed description of the proposed framework including the encoder and the decoder are given in Section III. Experimental results are presented in Section IV. Finally, concluding remarks are given in Section V.

\section{Principle of Compressed Sensing based Video Coding}

Compressed sensing is a signal processing technique that enables sub Nyquist sampling rate, with little or no drop in reconstruction quality, provided the signal $\bf{x}$ is sparse in some domain (DWT, DCT etc.). \\
Let $\bf{x} \in \mathcal{R}^{N}$ denote a real and discrete-time signal and $\bf{s} \in \mathcal{R}^{N}$ its transform domain ($\bf{\Psi}$) representation, i.e.,
\begin{equation}\label{eq1}
\bf{x} = \bf{\Psi} \bf{s} = \sum\limits_{i = 1}^N {\bf{\Psi}_i}{{s_i}}
\end{equation}
Let $\textbf{y}=\bf{\Phi}\textbf{x}=\bf{\Phi}\bf{\Psi}\textbf{s} \in \mathbb{R}^{M}$
represent a set of $M$ linear projections of $\bf{s}$. In compressed sensing, $\bf{s}$ is then reconstructed from a small set of linear projections $(M \ll N)$ by solving the following optimization problem,
\begin{equation}\label{eq2}
\min_{\bf{s}} \|\bf{s}\|_{0} \hspace{5mm} subject \hspace{5mm} to \hspace{5mm}   \bf{\Phi}\bf{\Psi}\textbf{s}=\bf{y}
\end{equation}
where $\|\bf{s}\|_{0}$ is called $\ell_0 $-norm and is defined as the number of non-zero components of  $\bf{s}$.\\

\par This is an under-determined linear equation problem with usually no unique solution. However, the signal $ \bf{s} $ can be recovered losslessly from $M \ge K$ measurements, if the sensing matrix $ \bf{\Phi} $ is designed in such a way that, it should preserve the geometry of the sparse signals and each of its $M \times K$ sub-matrices possesses full rank. This property is called Restricted Isometry Property (RIP) \cite{CS_12}\cite{CS_13} and mathematically, it ensures that $\|\bf{x}_1- \bf{x}_2\|_2 \approx\|\bf{\Phi} \bf{x}_1- \bf{\Phi} \bf{x}_2\|_2$, where $\|\bf{y}\|_2$ represents the  $ \ell_2 $- norm of the vector $\bf{y} $. It has been observed that the random matrices drawn from independent and identically distributed (i.i.d.) Gaussian or Bernoulli distributions satisfy the RIP property with high probability\cite{CS_13}.
\par
The problem of signal recovery from CS measurements is very well studied in the recent years and there exists a host of greedy algorithms that have been proposed such as Orthogonal Matching Pursuit (OMP) \cite{26}-\cite{28}, Iterative Hard-Thresholding (IHT) \cite{IHT}, Iterative Soft-Thresholding (IST) \cite{IST}. Approximate Message Passing (AMP) algorithm \cite{dono_1},\cite{dono_2},which  possesses  similar structure as that of IHT/IST, exhibits superior performance in terms of convergence rate and measurements trade-off. 

\begin{figure}
\centering
\includegraphics [height=40mm,width=80mm]{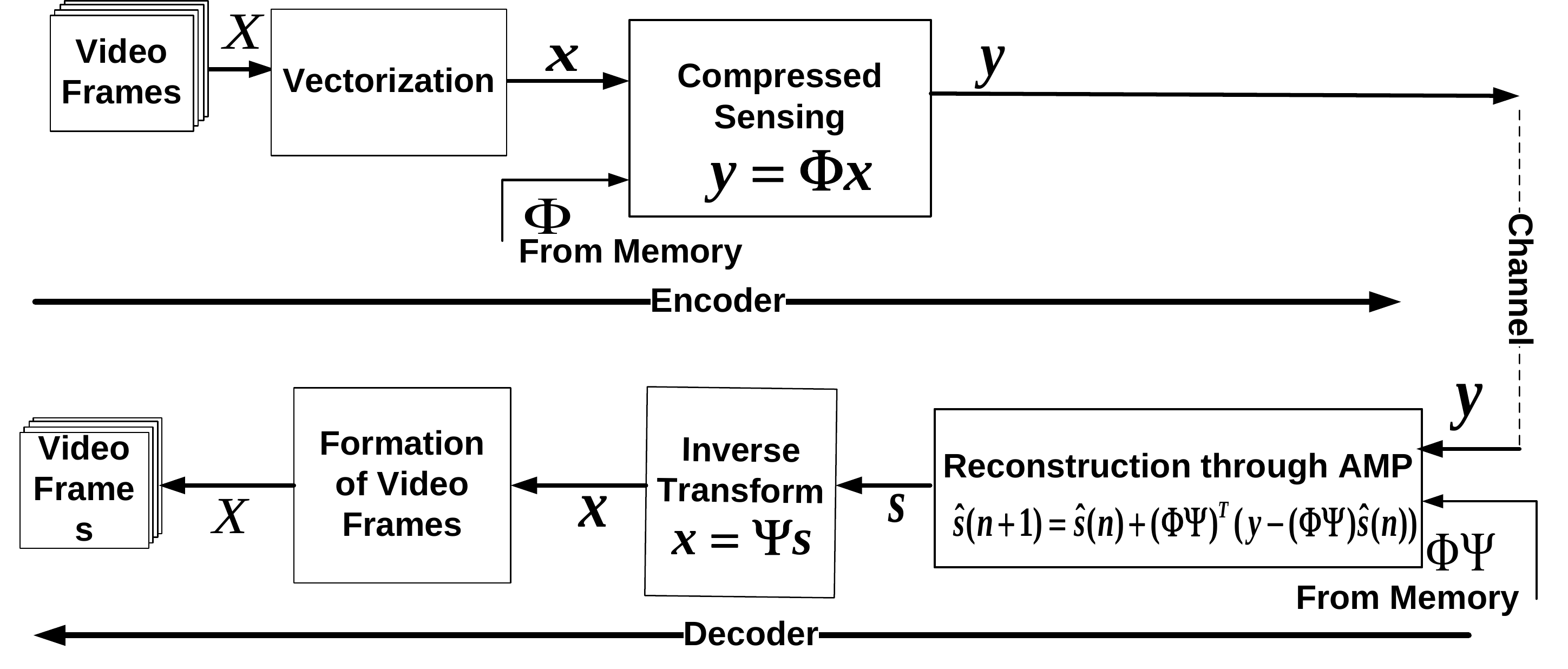}
\caption {Conventional CS based Video Codec}
\label{blockdia_1}
\end{figure}

Fig.~\ref{blockdia_1} represents the block diagram of conventional CS based video coding system, in which ${\bf X}$ represents the set of original video frames and $\bf{x}$ is the vectorization of X (i.e. $\textit{vec}(X)={\bf x}$). At the encoder, formation of the measurements ${\bf y}$ is done by multiplying the original signal ${\bf x}$ with sensing matrix ${\bf \Phi}$ i.e. $\bf{y} = \bf{\Phi} \bf{x}$, provided $\bf{x}$ is sparse in transform domain $\bf{\Psi}$. 

At the decoder, one solves the optimization problem \eqref{eq2} from the received measurements ${\bf y}$, by applying the AMP algorithm iteratively. Thereby one can reconstruct the transformed coefficient vector  (which is generally sparse) i.e., ${\bf s}$  by using the following equation:
\begin{equation}\label{eq3}
{\hat s{(n + 1)}} = {\hat s{(n)}} + {(\Phi\Psi) ^T}(y - \Phi\Psi {\hat s{(n)}})\hspace{3mm}n = 0, 1, .....Itr
\end{equation}
Finally, the original signal $\bf{x}$ is obtained from $\bf{s}$ by applying the inverse transform $ \bf{x}= \bf{\Psi} \bf{s}$.

Though the product $\bf{\Phi} \bf{\Psi}$ can be pre-computed and loaded into the memory, each iteration of \eqref{eq3} requires $2*M* N$ multiplications since $ \bf{\Phi} \bf{\Psi}$ is no longer a Bernoulli matrix whose elements are only $\pm 1$. Since these reconstruction algorithms are iterative by nature, a large number of iterations are required to ensure converge, which drastically increases overall complexity of the decoder. Hence, it is extremely difficult to realize the algorithm by VLSI implementation.  

\section{Proposed Framework for CS based Scalable Video Coding}
\begin{figure}
\centering
\includegraphics [height=80mm,width=90mm]{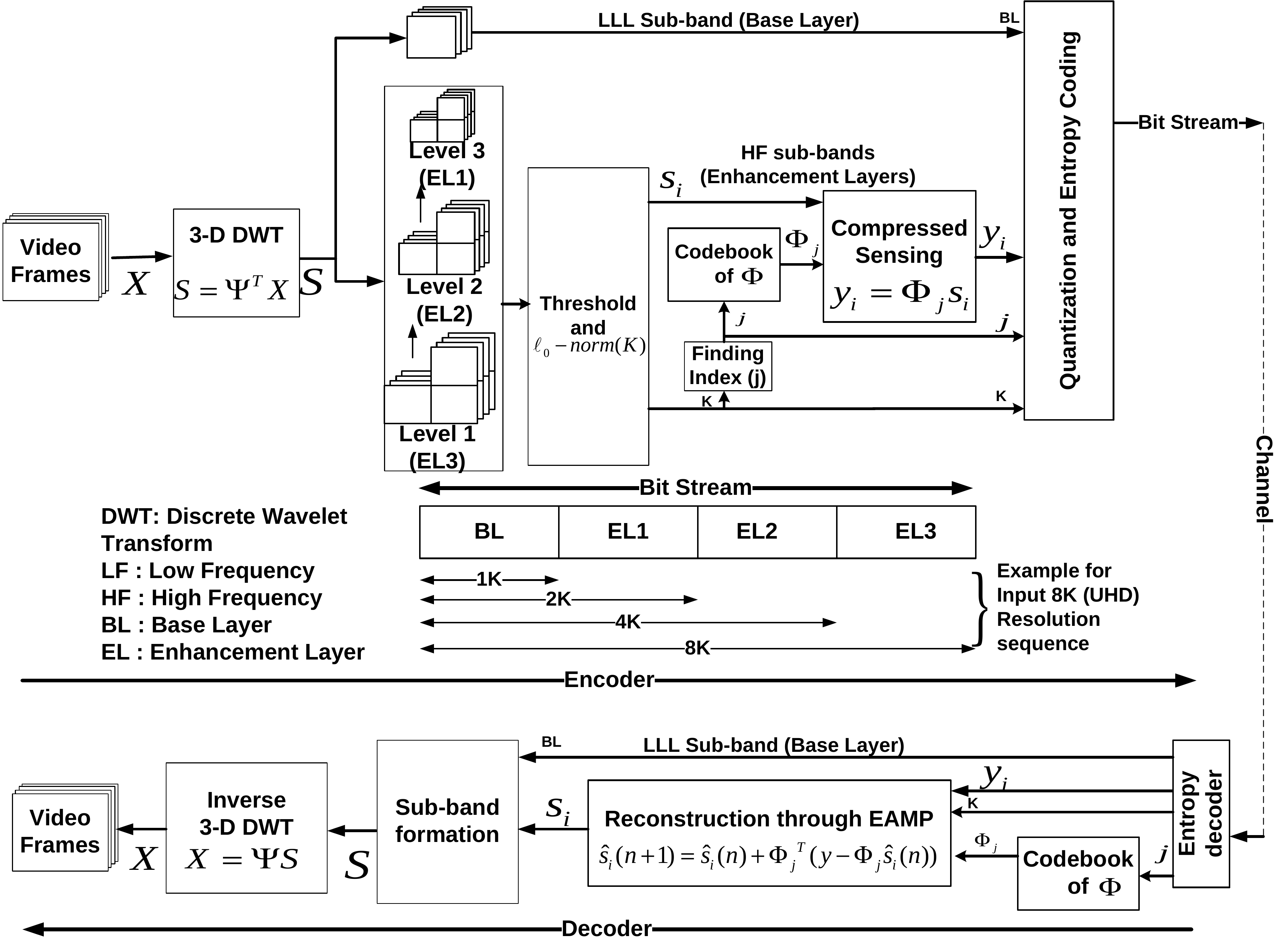}
\caption {Proposed CS based Scalable Video Codec}
\label{blockdia_2}
\end{figure}

Fig.~\ref{blockdia_2} shows the detailed block diagram of the proposed compressed sensing based scalable video encoder and decoder.

\subsection{CS based Video Encoder}
\par
At the first stage of the encoder, 3-D DWT has been applied on the input video frames.  Among all the sub-bands of the 3-D DWT, the LLL sub-bands (LL sub-bands of the L-frames), which do not possess any sparsity, are quantized and entropy coded directly without applying any CS techniques. The LLL sub-band is considered as the base layer in SVC domain, because it represents the image with all of its significant information in low resolution.  All the other sub-bands (3-D wavelet coefficients except the LLL sub-band) exhibit approximate sparsity (having coefficients with negligible values) and hard thresholding (i.e. the coefficient is considered to be zero if its value is less than a given threshold (T)) followed by compressed sensing is applied on them. CS measurements are quantized and then entropy coded before their  transmission from the encoder. The entropy coder codes the sample by Golomb Rice Coding (GRC) or Adjusted Binary Coding (ABC) based on the statistics obtained from the context modeler. The binary samples are further compressed by run length coding,  thus yielding the binary bits.

\subsubsection{3-D DWT and Scalability}
\par 
3-D DWT has been performed by combining spatial (2-D) and temporal (1-D) transforms. After applying 2-D DWT on each frame of the group of frames (GOF), the resulting DWT coefficients constitute the 4 sub-bands for each frame, viz. LL, HL, LH and HH sub-bands. When 1-D DWT in temporal direction is applied on the resultant spatial DWT coefficients, half of the frames become low frequency frames (L-frames), while the remaining half constitute the high frequency frames (H-frames). 

\par Inherent scalable feature of the DWT is used in the proposed framework to provide the spatial, temporal and quality scalabilities.  For an input video sequence with 8K resolution, Fig.~\ref{blockdia_2} depicts the formation of the base layer and the three enhancement layers. The decoder can acquire the necessary bit-streams starting from the last level (BL) with 1K resolution up to the first level (EL-3) with 8K resolution depending on the channel bandwidth and the display capacity of the user terminal. For different frame rates, one may have to receive every frame or drop some of the frames in order to match the network transmission condition.
\begin{figure}
\centering
\includegraphics [height=90mm,width=80mm]{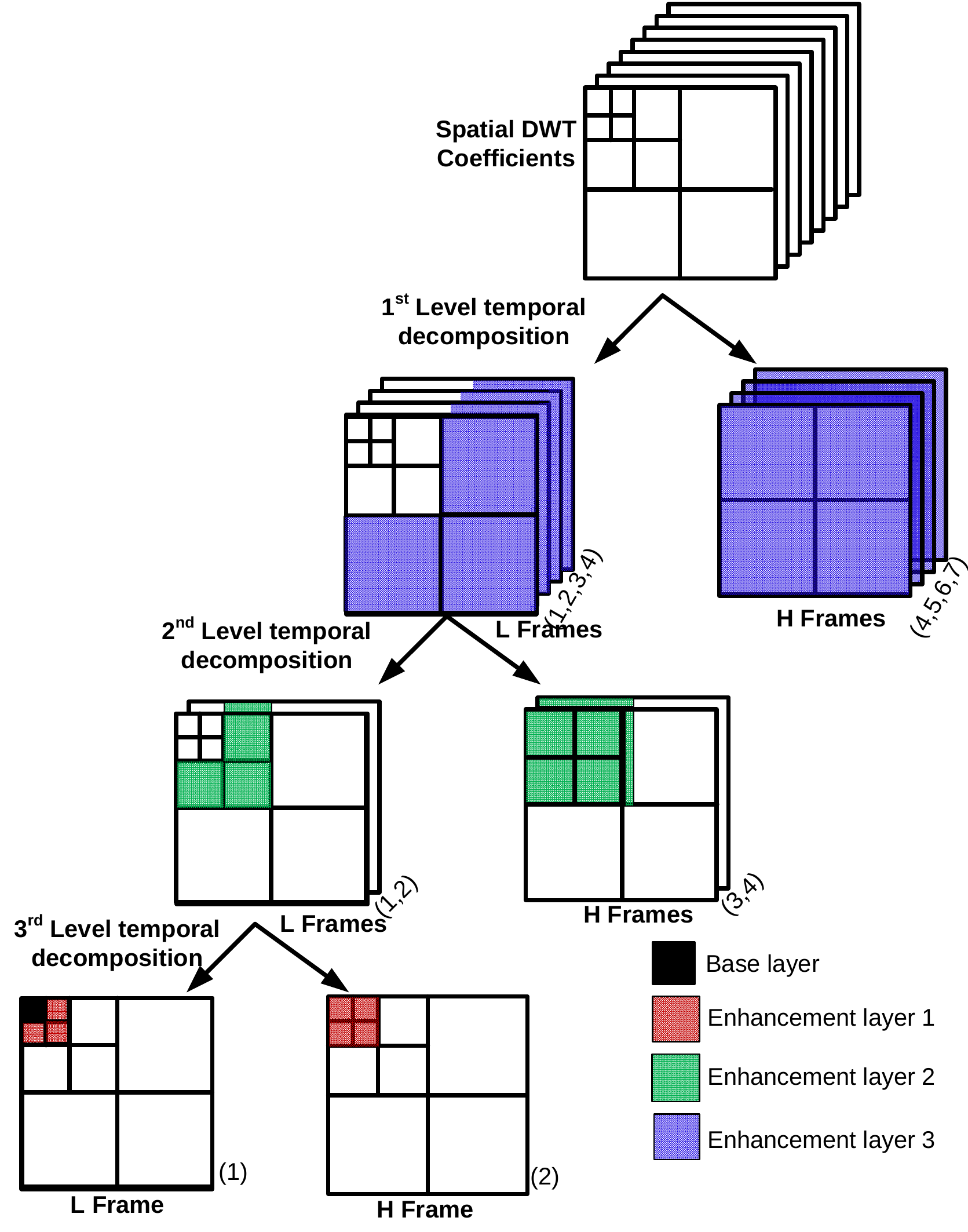}
\caption {Different levels of Spatial and Temporal scalability.}
\label{svc_13}
\end{figure}
\par LL sub-band derived from the highest level of decomposition is treated as the base-layer for the spatial scalable video bit-stream and the other sub-bands become the enhancement layers, thereby achieving spatial scalability. Temporal scalability is used to reduce the video frame rate and it is a common approach in cases where a limited channel bandwidth is available. Utilizing the multi-resolution property of wavelet transform along the temporal axis helps obtain temporal scalability.  Fig.~\ref{svc_13} depicts the temporal scalability process in the proposed framework, for a GOF of size 8.

\subsubsection{Adaptive Measurement Scheme}

Conventional CS based video encoders \cite{CS_10}-\cite{CS_plonka} find the CS measurements ($ \textbf{y} $) by considering the complete GOF or entire frame as a input vector $ \textbf{x} $ of size N$ \times $1, which is multiplied with the sensing matrix $ \bf{\Phi }$ of size M$ \times $N. It maintains a constant value of M for all the input vectors irrespective of their sparsity, which may lead to a decrease in compression ratio and increases the complexity for larger values of N.  For example, consider an HD video sequence with GOF of 8, for which the size of the input vector $\bf{(x)}$ becomes 1920$ \times $1080$ \times $8 = 16588800 = N. To get the CS measurements of size M$ \times $1, it is required to multiply two matrices of size (M$ \times $N)  and (N $\times $1), which requires a huge number of multipliers and a large amount of memory for  hardware implementation, which is impractical. The proposed framework addresses this problem by using an adaptive measurement scheme.

\begin{figure}
\centering
\includegraphics [height=60mm,width=100mm]{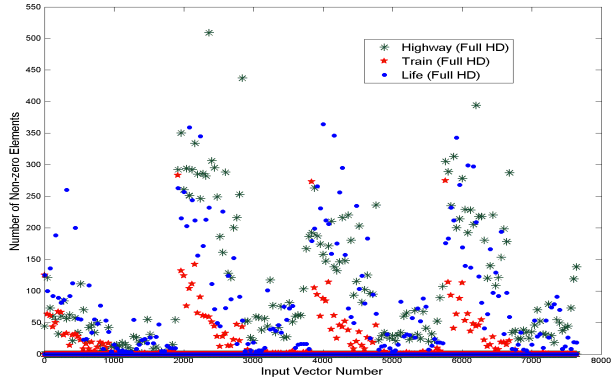}
\caption {Measurement of $ \ell_0 $-norm (K) of input vectors for different input sequences  }
\label{CS-vector}
\end{figure}
The following steps are involved in the adaptive measurement scheme, which is concerned with building a codebook common to the encoder and the decoder.
\begin{enumerate}
\item Form the input vector $\bf{(s_i)}$ for each high frequency sub-band with size N$ \times $1
\item Find the non-zero coefficients K ($ \ell_0 $-norm) in each input vector $\bf{(s_i)}$
\item find the index $ j $ of the code book based on range of $K$
\item Select the sensing matrix $ \bf{\Phi_{j}} $ from the codebook of Bernoulli matrices $ \bf{\Phi}$
\item Calculate the measurement vector $\bf{y_i = \Phi_{j} s_i}$
\end{enumerate} 
Formation of the input vector $\bf{(s_i)}$ and CS measurement vector ($\bf{y_i}$) is initiated from the last (third) level of temporal decomposition. At the third (as GOF is considered as 8) level of decomposition shown in Fig. \ref{svc_13}, top left corner block (represented with black color box in Fig. \ref{svc_13}) of the L-frame indicates the lowest frequency sub band (base layer) and is directly encoded after quantization without any CS process. The remaining 7 sub-bands in the third level spatial DWT (three from L frame and four from H frame, which are represented by red-colored box in Fig. \ref{svc_13}), are considered for the calculation of CS measurements sub-band wise. The size of each sub-band in level $l$ is W/$2^{l} \times $ H/$2^{l}$. To apply the CS on each sub-band, we convert the multiple columns into a single column vector (input vector $\bf{(s_i)}$) of dimension N$ \times $1, where N = $2^{n} \times $ H/$2^{l}$, n being a positive integer. The actual value of n depends on the frame size and level $ l $ of decomposition. The number of columns for each input vector $\bf{(s_i)}$ is equal to N/(H/$2^{l}$). The size of the randomly generated Bernoulli matrix  $ (\bf{\Phi_j}) $ is M$ \times $N, ($M \ge cK\log (N/K)$ for some small constant c). The range of values of  $ K $ ($ \ell_0 $-norm) for different video sequences is displayed in Fig. \ref{CS-vector}. This range varies from 0 to 510 for $ N $ = $2^{i} \times $H, where i is 1 for input video of HD resolution (1920$ \times  1080 $), 2 for 512 $\times $512 resolution and 3 for CIF resolution (352$ \times$  288). Based on the above observations, the codebook had been prepared with 16 entries, each entry (represented with index $ j $) has one sensing matrix $ (\bf{\Phi_j})$ with predefined M based on range of K (vide Table \ref{rangeofK}), and the same codebook is preserved at the decoder. The $ \ell_0 $-norm (K) and index $ j $ of the codebook are transmitted with the measurement vector $ \bf{y_i}$, to identify the actual Bernoulli matrix which needs to be used at the decoder. The $ \ell_0 $-norm (K) of each input vector $\bf{s_i}$ of length N is calculated and the Bernoulli matrix $ \bf{\Phi_{j}} $ is picked up from the codebook and the measurement vector $\bf{y_i = \Phi_{j} s_i}$ is calculated. This procedure is followed for all the $3^{rd}$ level spatial high frequency sub-bands of $3^{rd}$ level temporal frames (red-colored sub-bands shown in Fig. \ref{svc_13}). Each measurement vector $ \bf{y_i} $ is quantized and then subjected to entropy coding. Subsequently, the resulting coded bit streams are transmitted through the channel. On forming the CS measurement vectors for the $3^{rd}$ level sub-bands, same procedure is performed for the $2^{nd}$ level sub-bands(shown as green-colored boxes in Fig. \ref{svc_13}), and finally for the $1^{st}$ level sub-bands(shown as blue-colored boxes in Fig. \ref{svc_13}).
\begin{table*}[]
\centering
\caption{Number of measurements (M) for different range of K values}
\label{rangeofK}
\scalebox{0.88}{\begin{tabular}{|c|c|c|c|c|c|c|c|c|}
\hline
Range of K & 0 & 0$<$K$\leq$10 & 10$<$K$\leq$20 & 20$<$K$\leq$50 & 50$<$K$\leq$100 & 100$<$K$\leq$150 & 150$<$K$\leq$200 & 200$<$K$\leq$250 \\\hline
Value of M & 0 & 50     & 130     & 240     & 370      & 470       & 650       & 780        \\ \hline
Index (j)     & 0 & 1      & 2       & 3       & 4        & 5         & 6         & 7             \\ \hline
\hline

Range of K& 250$<$K$\leq$300 & 300$<$K$\leq$350 & 350$<$K$\leq$400 & 400$<$K$\leq$450 & 450$<$K$\leq$500 & 500$<$K$\leq$550 & 550$<$K$\leq$600 & 600$<$K \\ \hline
Value of M& 920       & 1080      & 1220      & 1400      & 1550      & 1700      & 1850      & 2000  \\ \hline
Index (j)& 8         & 9         & 10        & 11        & 12        & 13        & 14        & 15\\ \hline
\end{tabular}}
\end{table*}
\subsection{CS based Video Decoder}
   As depicted in Fig.~\ref{blockdia_2}, the decoder receives the coded bits and decodes them using an entropy decoder. The present work proposes an enhanced AMP  (EAMP) algorithm to recover the sub-band coefficients from the measurement vector.  Fig. \ref{CS_graph} shows the proposed EAMP algorithm, which is derived from AMP and IHT algorithms, provides the better performance compared to the existing iterative algorithms.

\subsubsection{Sparse Recovery using EAMP algorithm}
\begin{figure}
\centering
\includegraphics [height=40mm,width=90mm]{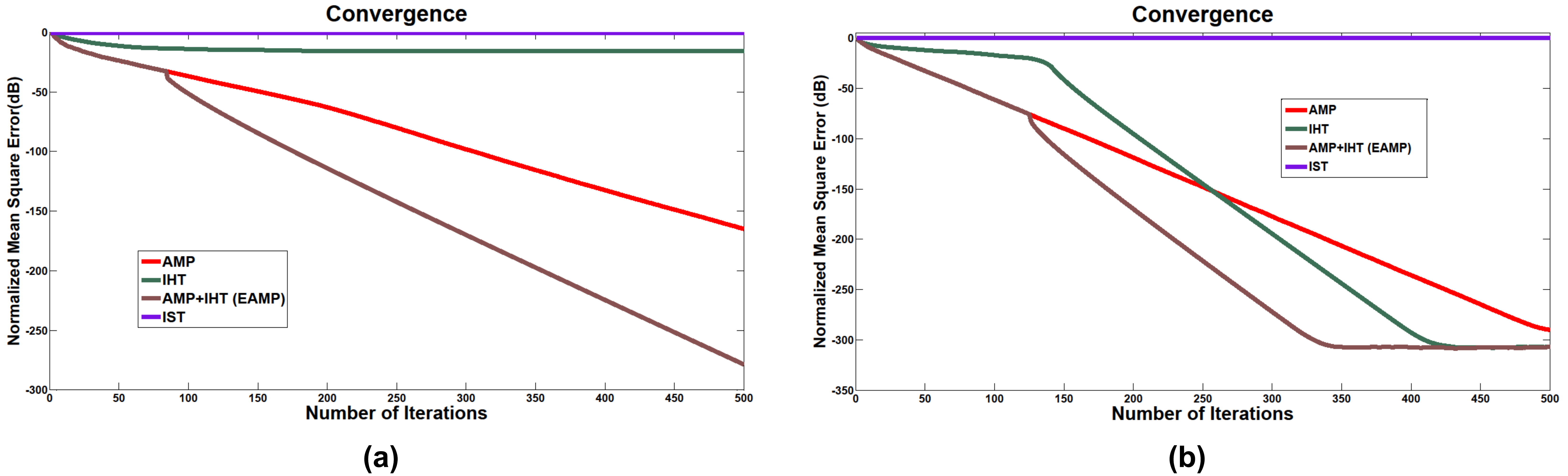}
\caption {Normalized mean square error of (a) $5^{th}$ input vector of `Life' sequence having K=176 (b) $10^{th}$ input vector of `foreman' sequence having K=82}
\label{CS_graph}
\end{figure}
The available gradient pursuit algorithms like Iterative Soft Thresholding (IST) \cite{IST} and Iterative Hard Thresholding (IHT) \cite{IHT} are less complex when compared to $\ell_{1}$-norm based algorithms. However, they perform inadequately with respect to performance measurement trade-off (\textit{viz.} reconstruction probability vs. number of measurements). The Approximate Message Passing (AMP) algorithm  \cite{dono_1},\cite{dono_2} has a structure similar to that of IST, with an extra term derived from the message passing algorithms. This significantly improves the performance measurement trade-off of AMP algorithm when compared to IST and IHT, with a minimal increase in complexity. Availability of prior information about the exact sparsity $K$ along with enough measurements allows IHT to provide better steady state Mean Square Difference (MSD) performance compared to AMP algorithm. Since the proposed scheme is involved with calculating $\ell_{0}-norm$ at the encoder and transmitting it to the decoder along with the measurements, this prior information can be used to enhance the MSD performance of AMP. 

Since the AMP algorithm can reconstruct the wavelet coefficients with very few measurements compared to IHT algorithm, the proposed EAMP algorithm proceeds as the AMP algorithm till the support set, which consists of indices of active coefficients,  is obtained. Then the EAMP algorithm follows the structure of IHT algorithm in order to achieve good MSD performance. The proposed EAMP algorithm, which owes its origin to both AMP and IHT algorithms in order to get the benefits of both, is shown as Algorithm 1. \\

\par The learning curves (Normalized Mean Square Error (NMSE) vs. Iterations) are depicted in Fig. \ref{CS_graph} (a) and (b) for two input vectors extracted from the 3-D DWT coefficients of the two video sequences.  Fig. \ref{CS_graph}(a) is drawn by taking the input vector number 5 of `Life' video sequence with length (N) 2160, measured  $ K $ ($ \ell_0 $-norm) is 176, number of measurements $ (M)$ considered for a range of  $150 < $ K $ < 200$ is 650. Table \ref{rangeofK} shows the values of M for different range of values of K. Fig. \ref{CS_graph} (b) is drawn by taking the input vector number 10 of `Foreman' video sequence with length 2048, measured  $ K $ is 82, and the number of measurements $(M)$ considered for a range of  $50 < $ K $ < 100$ is 370. An inspection of Fig. \ref{CS_graph} (a) and (b) clearly shows the superiority of the proposed EAMP over the AMP \cite{dono_1}\cite{dono_2}, IHT \cite{IHT} and (IST) \cite{IST}.\\
  
\begin{algorithm}
\KwIn{$y_i, j, K $}  \tcp{$y_i$:Measurement Vector, j: Index}
\tcp{K:$\ell_{0}$-norm}
\KwOut{$ \bf{\hat{s_i}}$} \tcp{$ \bf{\hat{s_i}}$:Reconstructed Input vector}
\caption{{\bf Enhanced Approximate Message Passing Algorithm (EAMP)}
\label{Algorithm}}
 $\bf{\Phi}  \gets codebook(j); $\\
 $N \gets sizeof(\bf{\Phi} ,2); $\\
 $M \gets sizeof(\bf{\Phi} ,1); $\\
 $ {\bf{\hat{s}}} \gets {nullset};  $\\
 $z \gets y_i; $\\
 $Iter \gets 400; $\\
\For{$i \gets 1$ \textbf{to} $Iter$}
{
	\If {(i $<$ Iter/4)}
	{
    $ \bf{\gamma} \gets  \bf{\hat{s_i}} + \bf{\Phi^T}*z;$  \\
     $temp \gets sort(abs({\bf{\gamma}}), `descend');$\\
     ${\bf{\delta}} \gets temp(M);$ \\
     $clear temp;$      \\
	
    $tmp \gets (abs({\bf{\gamma}})-{\bf{\delta}});$\\
    $tmp \gets (tmp+abs(tmp))/2;$\\
    ${\bf{\hat{s_i}}}    \gets sign({\bf{\gamma}}).*tmp;$\\
    $ampterm \gets (z/n)*sum(({\bf{\gamma}} > {\bf{\delta}}) + ({\bf{\gamma}}  < -{\bf{\delta}}));$\\
    $z \gets y_i - {\bf{\Phi}} *{\bf{\hat{s_i}}}  + ampterm;$\\
    }
    \Else
    {
    $    \bf{\hat{s_i}}  \gets \bf{\hat{s_i}}  +\bf{\Phi^T}*z;$\\
    $[value, Addr] \gets sort(abs({\bf{\hat{s_i}} }),`descend'); $\\
    \For{$\ell \gets K+1$ \textbf{to} $N$}
    {
    ${\bf{\hat{s_i}}} (Addr(\ell,1))\gets 0;$\\
    $ampterm \gets 0;$\\
    }
     $z \gets y_i - {\bf{\Phi}} *{\bf{\hat{s_i}}}  + ampterm;$\\
	}	
}
\Return{$\bf{\hat{s_i}} $}\\
\end{algorithm}

\par Algorithm \ref{Algorithm} shows the pseudo-code for reconstruction of input vector $ \bf{s_i}$ from the measurement vector $ \bf{y_i} $ through the proposed EAMP algorithm. The encoder sends the measurement vector $ \bf{y_i} $, the index $ j $ of the Bernoulli matrix in the codebook and the $ \ell_0 $-norm (K) of the input vector $ \bf{s_i}$. Here, the codebook consists of 16 Bernoulli matrices and the index indicates the Bernoulli matrix which has been used at the encoder. Recall that same Bernoulli matrix is utilized at the decoder. The decoder estimates the sparse signal $ \bf{\hat{s}} $ which is very similar to $\bf{s} $ by executing the EAMP algorithm. In the next stage, $ \bf{\hat{s}} $ is rearranged in such a way that it forms the wavelet sub-bands back. This process is opposite in nature to what is performed at the encoder end. At the encoder, the multicolumns of the sub-bands are arranged in a single column, whereas at the decoder it is the other way. On forming this arrangement, the inverse wavelet transform is applied and the output constitutes a video sequence which is almost an exact replica of the input video sequence. 

\section{Simulation Results and Performance Analysis}
 The proposed framework has been implemented in MATLAB. In the process of 3-D DWT, lifting based 9/7 biorthogonal filters \cite{SVC_21} are used for spatial DWT (2-D),  and Haar wavelets filters are used for temporal direction (1-D) to reduce the on-chip memory in hardware implementation.  As described in section III.A, after performing 3-levels of wavelet decomposition, the resultant high frequency wavelet bands are sent to CS based adaptive measurement scheme. The resultant CS measurements are passed through the entropy coder (GRC/ABC). Note that for comparison, the percentage of measurements is calculated before the entropy coding, and compression ratio denotes the ratio of the total number of bits in input frame and the number of bits after the entropy coding. 

The entropy decoder at the decoder extracts the received index j, $ \ell_0 $-norm (K) of the input vector and the measurement vector $ \bf{y_i}$. As shown fig. 2 the EAMP algorithm is executed on the output of the entropy decoder. As a result, the approximate input vector $ \hat{s_i} $ is recovered from $y_i$. On recovery of all sub-bands of GOF, inverse DWT is applied on the recovered sub-bands, and thereby the output video sequence is obtained.

\begin{table*}[]
\centering
\caption{Reconstruction Quality of the proposed framework for various resolutions}
\label{table1}
\resizebox{7 in}{!}{%
\begin{tabular}{|c|c|c|c|c|c|c|c|c|c|c|}
\hline
\multirow{2}{*}{Video Clip} & Input & Threshold & Percentage of & \multicolumn{4}{c|}{Compression Ratio}      & PSNR for         & PSNR for 50\% of & PSNR for 25\% of \\ \cline{5-8}
                            &                 Resolution            & value (T) & Measurements  & CS      & CS + HE & CS + EC & CS + EC + RLE & Input resolution & Input resolution & Input resolution \\ \hline

Viplane                     & 256$\times$256                     & \multirow{9}{*}{$ T =1$}    & 19.60   & 5.1     & 8.4     & 10.2    & 21.16         & 36.1597          & 38.1607          & 43.5518          \\ \cline{1-2} \cline{4-11}
News                        & cif                         &                                & 31.91   & 3.133   & 5.1     & 6.7     & 12.1          & 46.4105          & 50.6762          & 56.8434          \\ \cline{1-2} \cline{4-11}
Container                   & cif                         &                                & 36.52    & 2.738   & 4.01    & 5.92    & 10.89         & 44.0341          & 48.9695          & 53.6423          \\ \cline{1-2} \cline{4-11}
Clock                       & 512$\times$512                     &                                & 20.43    & 4.8928  & 9.1     & 16.2    & 30.56         & 41.8892          & 49.9835          & 54.0726          \\ \cline{1-2} \cline{4-11}
Cyclone                     & 512$\times$512                     &                                & 28.38   & 3.5233  & 5.6     & 7.8     & 14.85         & 32.3118          & 38.2051          & 44.7179          \\ \cline{1-2} \cline{4-11}
Train\_HD                   & Full HD                     &                                & 27.70   & 3.61    & 5.7     & 7.6     & 14            & 42.4217          & 47.6274          & 51.2196          \\ \cline{1-2} \cline{4-11}
Elephantdream               & Full HD                     &                                & 29.68   & 3.3691  & 4.3     & 6.9     & 13.6          & 35.3581          & 43.768           & 49.5443          \\ \cline{1-2} \cline{4-11}
Highway\_HD                 & Full HD                     &                                & 26.80   & 3.7313  & 6.1     & 8.2     & 14.6          & 38.2862          & 45.5534          & 50.2317          \\ \cline{1-2} \cline{4-11}
Life                        & Full HD                     &                                & 34.01   & 2.94    & 4.1     & 6.8     & 11.76         & 45.2881          & 49.7547          & 53.5673          \\ \hline
Viplane                     & 256$\times$256                     & \multirow{9}{*}{$ T = 1.6$} & 6.10   & 16.3896 & 29.57   & 38.96   & 63.78         & 27.3607          & 28.1533          & 29.0984          \\ \cline{1-2} \cline{4-11}
News                        & cif                         &                                & 6.22   & 16.07   & 30.1    & 39.97   & 63.35         & 36.6489          & 41.8732          & 47.572           \\ \cline{1-2} \cline{4-11}
Container                   & cif                         &                                & 6.19   & 16.155  & 29.8    & 40.2    & 64.145        & 33.6307          & 38.512           & 43.237           \\ \cline{1-2} \cline{4-11}
Clock                       & 512$\times$512                     &                                & 4.96   & 20.1347 & 38.2    & 49.4    & 78.6          & 34.8619          & 38.4467          & 43.4998          \\ \cline{1-2} \cline{4-11}
Cyclone                     & 512$\times$512                     &                                & 4.53   & 22.0447 & 43.89   & 56.34   & 81.2          & 26.7005          & 28.5599          & 31.9933          \\ \cline{1-2} \cline{4-11}
Train\_HD                   & Full HD                     &                                & 4.66    & 21.4506 & 30.7    & 43.4728 & 78.0475       & 32.5365          & 34.7643          & 39.7336          \\ \cline{1-2} \cline{4-11}
Elephantdream               & Full HD                     &                                & 10.60   & 9.4318  & 14.5    & 19.859  & 39.4261       & 30.0076          & 38.212           & 43.109           \\ \cline{1-2} \cline{4-11}
Highway\_HD                 & Full HD                     &                                & 8.74   & 11.441  & 17.3    & 23.9702 & 46.8251       & 31.0496          & 32.948           & 36.1647          \\ \cline{1-2} \cline{4-11}
Life                        & Full HD                     &                                & 4.62   & 21.6    & 31.7    & 37.6    & 79.4          & 34.0895          & 37.5055          & 41.7916          \\ \hline
\multicolumn{11}{l}{CS: Compression Ratio with Compressed Sensing (Proposed)}                                                                                                                                                                                                                                                                                                                                                        \\
\multicolumn{11}{l}{CS+HE :Compression Ratio with CS and Traditional Huffman coding}                                                                                                                                                                                                                                                                                                                     \\
\multicolumn{11}{l}{CS+EC : Compression Ratio with CS and  GRC/ABC (Proposed Entropy Coding)}                                                                                                                                                                                                                                                                                                            \\
\multicolumn{11}{l}{CS+EC+RLE : Compression Ratio with CS and  GRC/ABC (Proposed Entropy Coding) + Run Length Encoding}
\end{tabular}}%
\end{table*}

\begin{table}[]
\centering
\caption{Performance comparison of proposed framework with TV-DCT \cite{31} }
\label{table_4}
\resizebox{\columnwidth}{!}{%
\begin{tabular}{|c|c|c|c|c|}
\hline
Video     & Compression & \multirow{2}{*}{Resolution} & \multicolumn{2}{c|}{PSNR}            \\ \cline{4-5}
                 Sequence          &           Ratio                   &                             & Proposed & Li \textit{et al.} \cite{31} \\ \hline
\multirow{3}{*}{Life}      & 35                           & 1080$\times$1920                   & 45.28    & 39.58                     \\ \cline{2-5}
                           & 9                            & 540$\times$960                     & 39.95    & 40.08                     \\ \cline{2-5}
                           & 2                            & 270$\times$480                     & 35.31    & 40.02                     \\ \hline
\multirow{3}{*}{Rush hour} & 35                           & 1080$\times$1920                   & 47.381   & 42.16                     \\ \cline{2-5}
                           & 9                            & 540$\times$960                     & 43.11    & 43.93                     \\ \cline{2-5}
                           & 2                            & 270$\times$480                     & 39.27    & 44.37                     \\ \hline
\end{tabular}}%
\end{table}
 
Extensive simulation has been performed on different test video sequences of sizes starting from over 256$\times$256, 352$\times$288, 512$\times$512 and 1920$\times$1080. The performance of the proposed scheme has been compared with the state of the art CS based video codecs \cite{CS_10}-\cite{CS_plonka} and the results are summarized as follows:\\
\begin{figure}[h!]
\centering
\includegraphics [height=40mm,width=70mm]{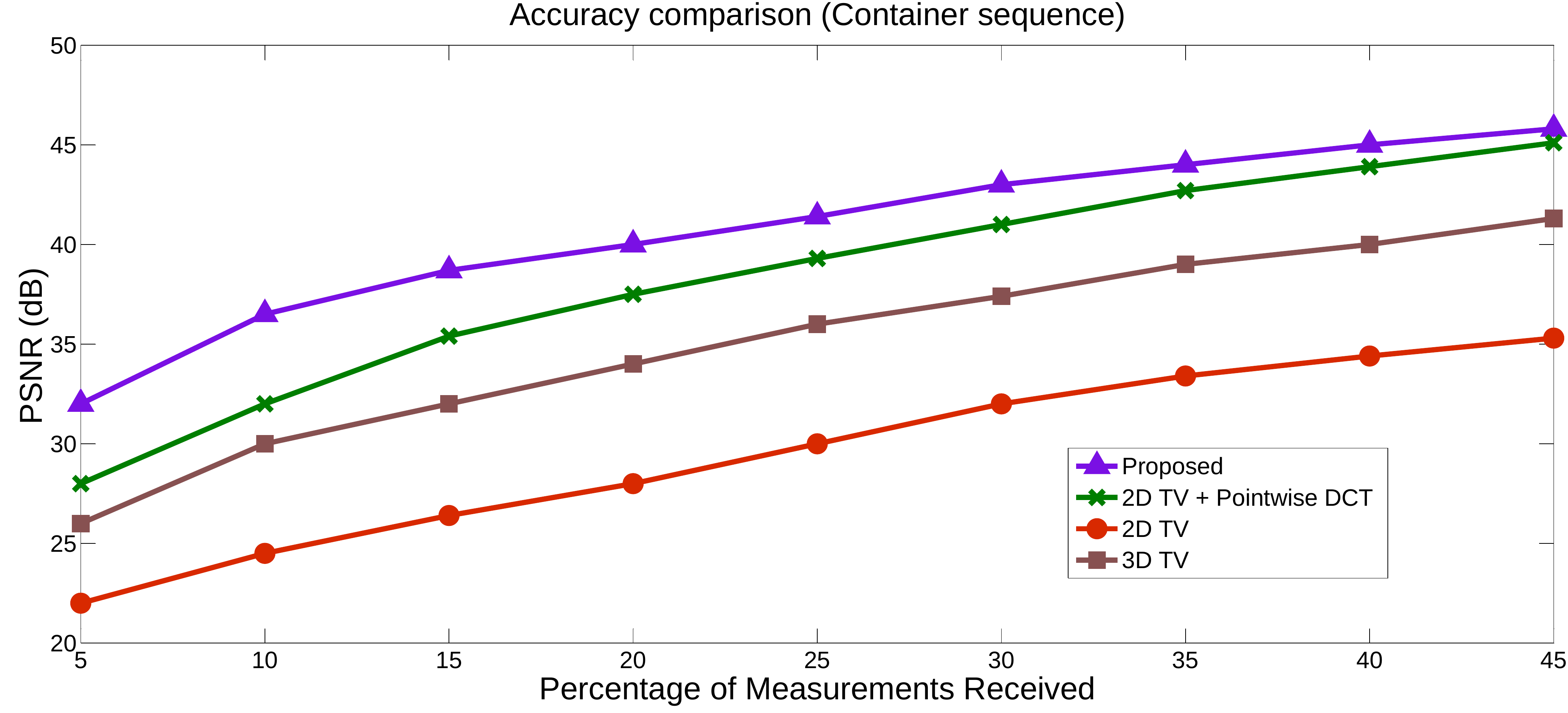}
\caption {Accuracy comparison of the proposed framework with existing methods for the \emph{`container'} video}
\label{result_plot1}
\end{figure}
\begin{enumerate}
\item Reconstruction quality (PSNR) vs. percentage of the measurements is plotted in Fig. \ref{result_plot1} for the container sequence having standard CIF resolution (352$\times$288). It can be observed that the proposed framework outperforms all the existing CS based methods  proposed by Li \textit{et al.} \cite{31}. \\
\begin{figure}[h!]
\centering
\includegraphics [height=60mm,width=90mm]{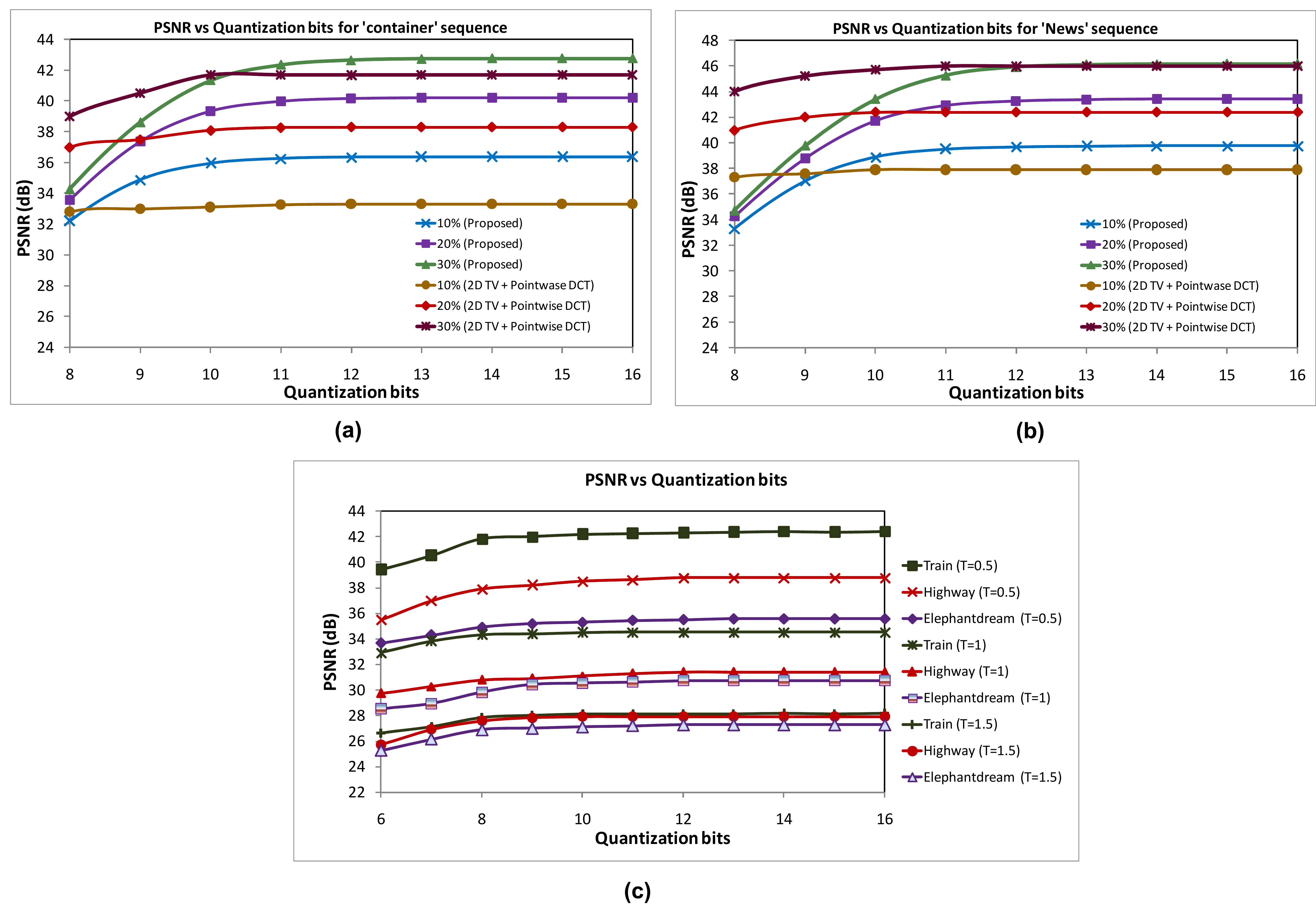}
\caption {Effect of quantization bits on reconstruction quality}
\label{result_plot2}
\end{figure}

\item  As quantization is performed on the CS measurements prior to their transmission at the encoder end, one should therefore naturally examine the reconstruction quality of the video at the decoder by varying the number of bits involved in the quantization process. Tests were conducted on `container.cif' and `news.cif' of CIF resolution, and results are compared with the 2D TV + Pointwise DCT proposed by Li \textit{et al.} \cite{31}. From Fig. \ref{result_plot2} (a) and (b), it is evident that the proposed method gives better performance or at least equal performance in the worst case compared to 2D TV + Pointwise DCT. We also performed the same tests on three videos of full HD resolution with varied threshold value T, and the results are plotted in Fig. \ref{result_plot2}(c). The number of measurements is reduced with an increase in threshold T, since coefficients less than T are considered as zero, which in turn ensures the reduction of K. From the results, it can be concluded that the reconstruction quality remains unaltered if the number of  quantization bits exceeds 10 per measurement. However, the quality increases with the number of measurements received.\\
\begin{figure}[h!]
\centering
\includegraphics [height=70mm,width=70mm]{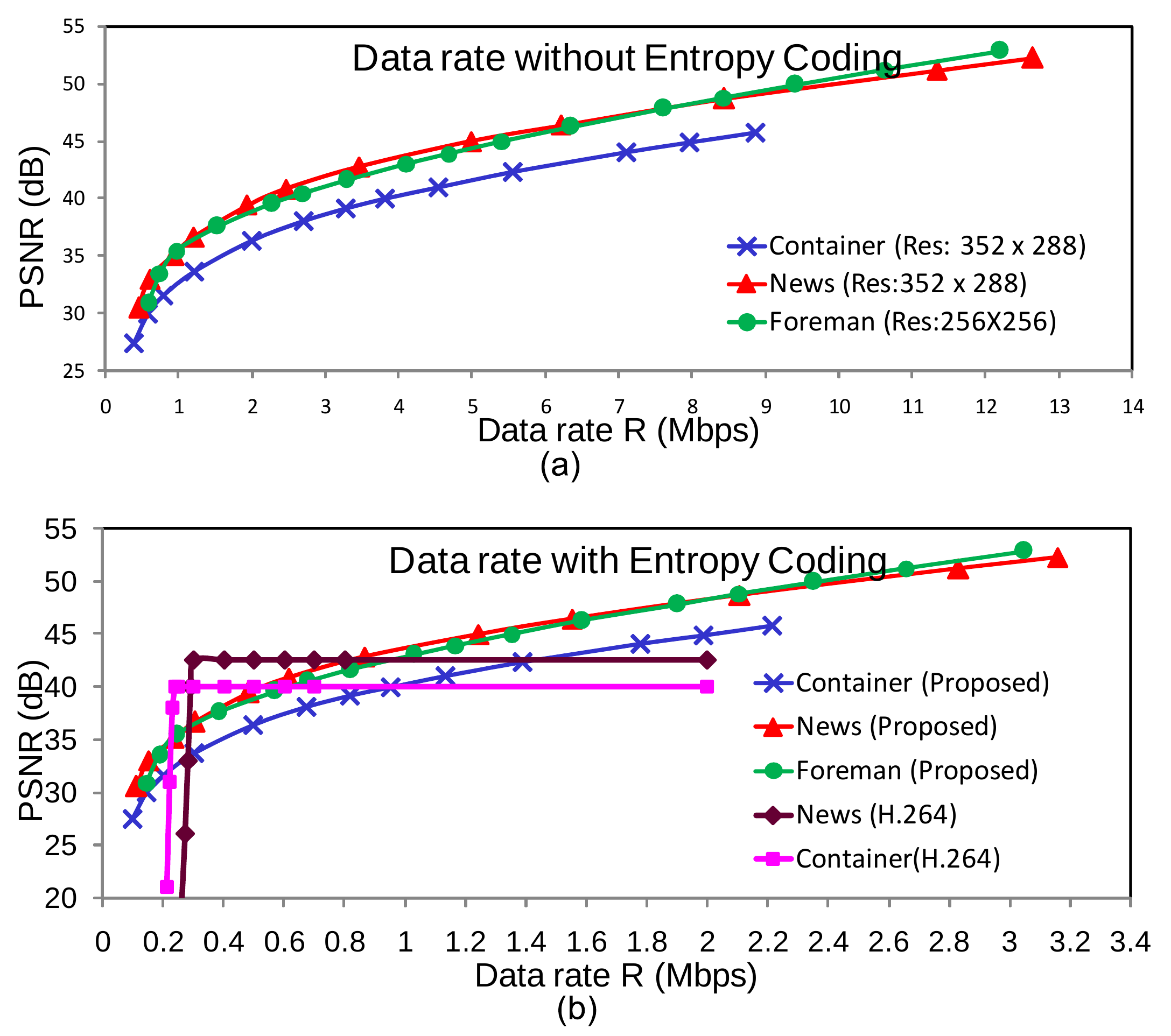}
\caption {Variation of PSNR with data rate for the `cif' resolution videos}
\label{result_datarate}
\end{figure}
\begin{figure}[h!]
\centering
\includegraphics [height=70mm,width=70mm]{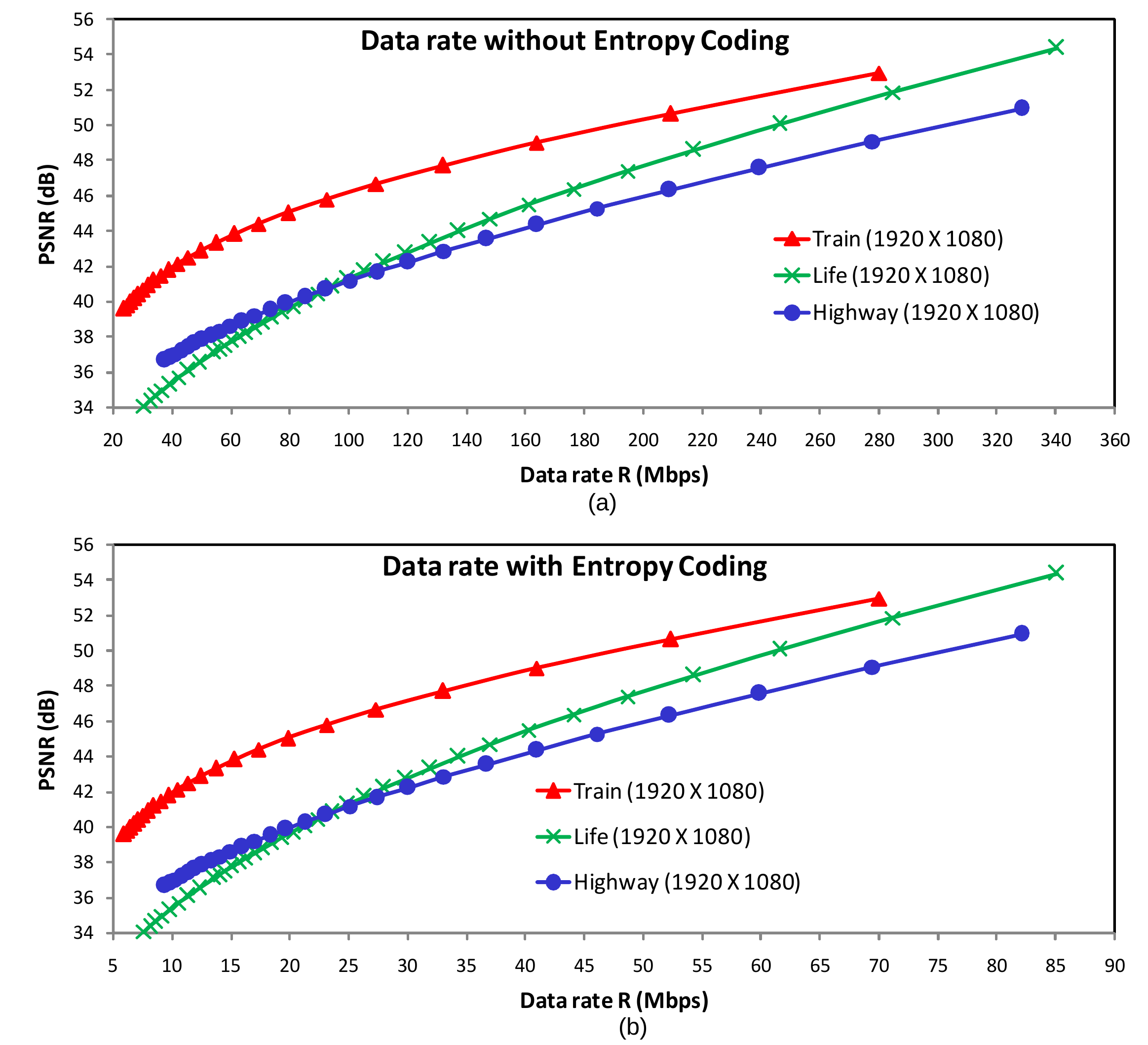}
\caption {Variation of PSNR with data rate for the `full HD' resolution videos}
\label{result_datarate_hd}
\end{figure}
\item  We verified the variation of PSNR as a function of received data rate R of the proposed CS based framework for the video sequences `foreman' (256$\times$256), `news.cif' and `container.cif' and the results are plotted for the system without and with entropy coding in Fig. \ref{result_datarate}(a) and (b) respectively. The same test has been performed on three full HD videos and results were plotted in Fig. \ref{result_datarate_hd}(a) and (b). The two plots shown in Fig. \ref{result_datarate} and  Fig. \ref{result_datarate_hd} demonstrate that PSNR increases with an increase in data rate R, where R represents the number of measurements received per second. Even if the received data rate is less than the designed data rate, the decoder is able to reconstruct the video with reasonable PSNR. The plots of Fig. 8 further establishes that the PSNR increases with an increase in the received data rate. Inspection of Fig. \ref{result_datarate}(b) indicates that an H.264 video codec suffers a sharp drop in video quality when the received data rate is below the specified minimum data rate $R_{d}$ and is constant even data rate is equal or much higher than the minimum data rate. The results for H.264 in Fig. \ref{result_datarate}(b) are sourced from Li \textit{et al.} in \cite{31}.\\

\item  We have investigated the performance of the proposed framework with the videos having different resolutions for the threshold values T=1 and T=1.6. Table \ref{table1} shows the different performance metrics viz. the percentage of measurements received without entropy coding, compression ratio (CR) provided with only CS module, CR with CS + traditional Huffman encoding, CR with CS + new GRC/ABC encoding, CR with CS + new GRC/ABC encoding + Run length encoding. Further Table \ref{table1} records the quality in terms of average PSNR in dB scale of the reconstructed frames with input resolution re-sized to 1/2 and $1/4^{th}$ of the original resolution.\\
\begin{figure}[h!]
\centering
\includegraphics [height=40mm,width=70mm]{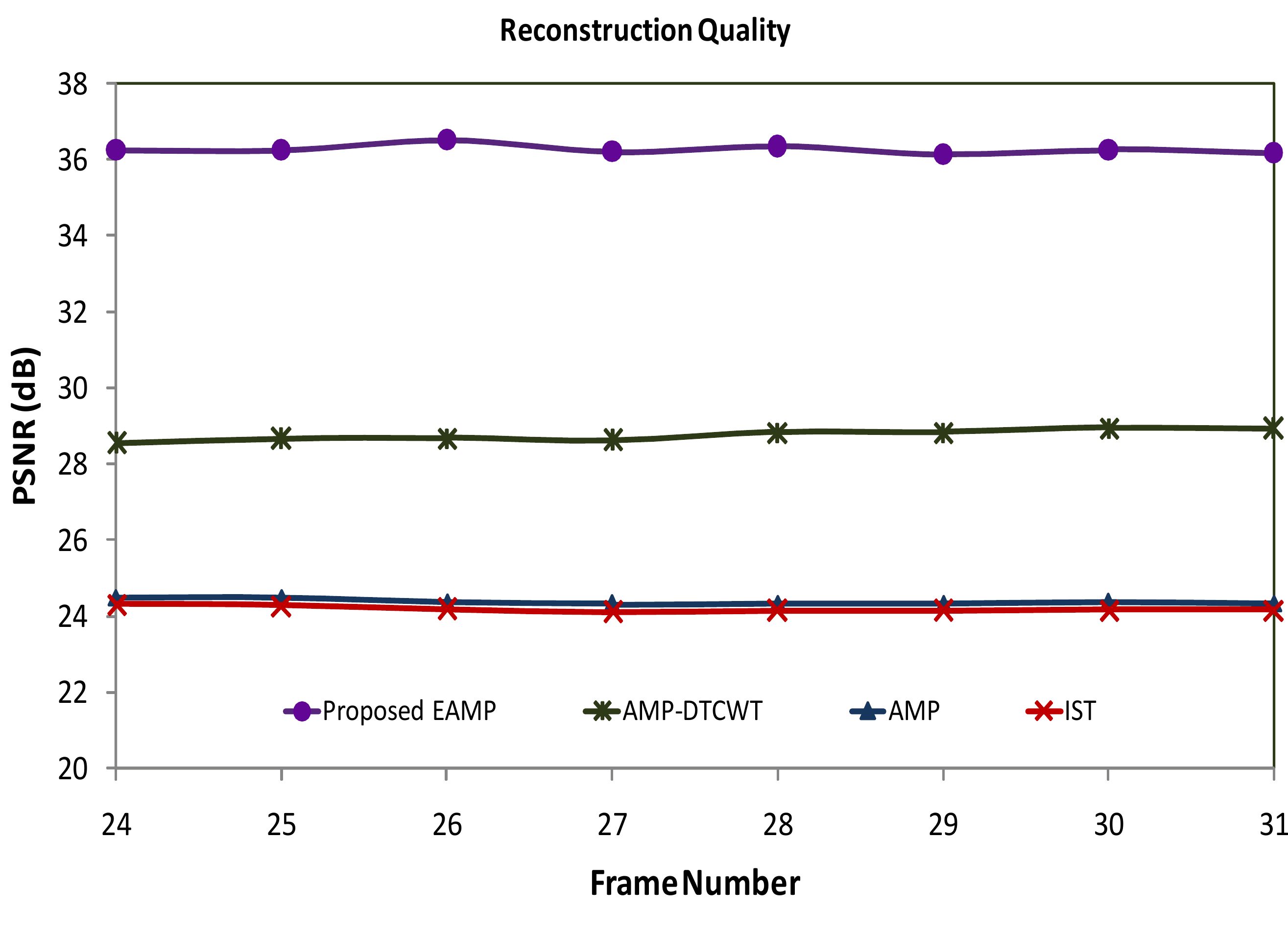}
\caption{Reconstruction quality (PSNR in $dB$) comparison of different recovery algorithms for `$Foreman$' video sequence}
\label{compare_plonka}
\end{figure}

\begin{figure*}
\centering
\includegraphics [height=140mm,width=150mm]{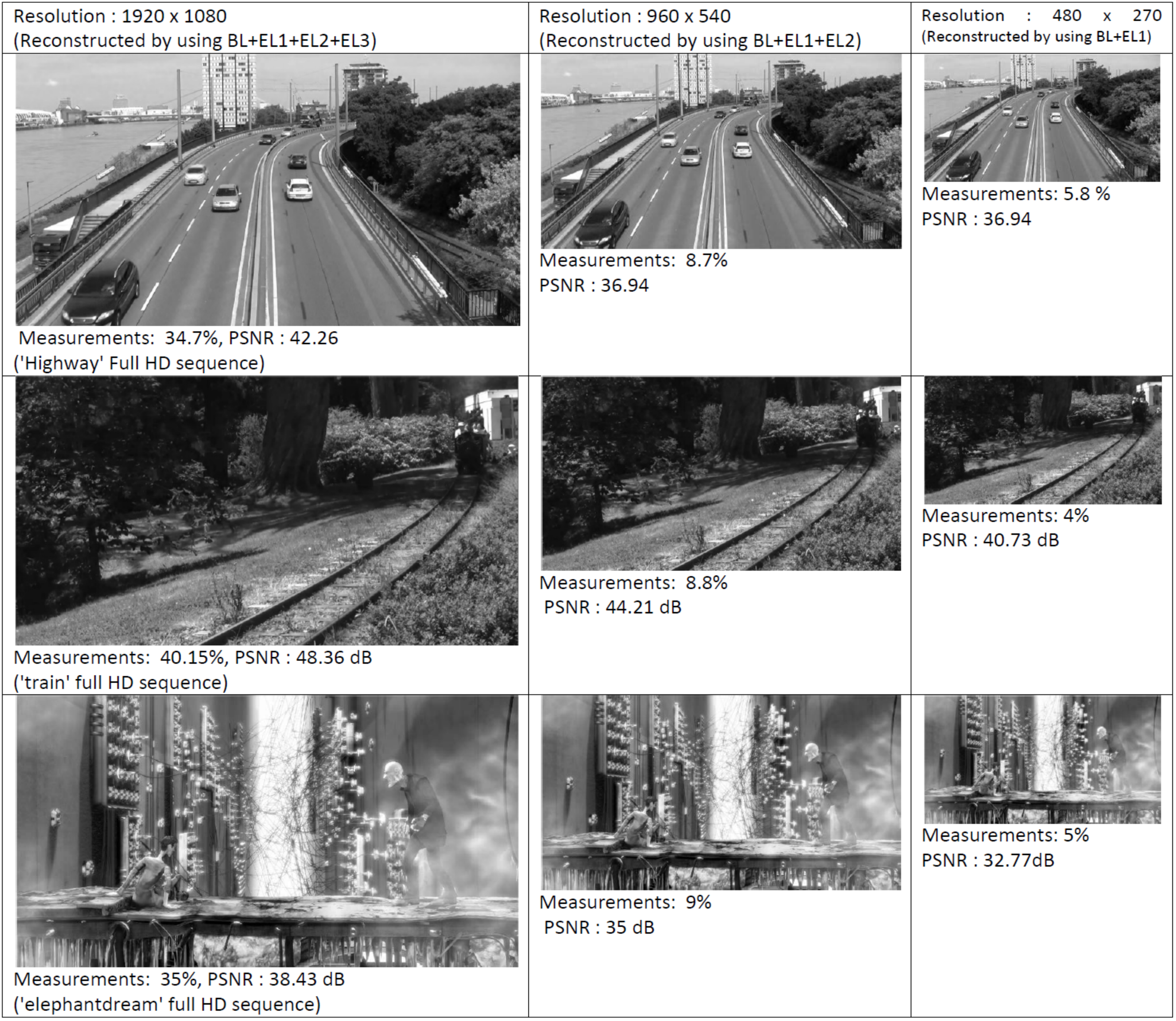}
\caption {Reconstructed frames for different levels obtained in the proposed framework}
\label{result_image}
\end{figure*}
\item  Fig. \ref{compare_plonka} depicts the performance of the proposed decoder \textit{vis-\`{a}-vis} that of the other existing recently proposed CS-based decoders \cite{CS_plonka} on the basis of PSNR (expressed in dB) for the frames 24-31 of the $ foreman $ sequence with 30$ \% $ measurements and 20 iterations. It indicates that the proposed framework with adaptive measurement scheme and EAMP as reconstruction algorithm produces better results compared to IST, AMP and AMP-DTCWT  \cite{CS_plonka} for the same number of measurements and iterations. This may be attributed to the feature of the proposed AMS procedure in calculating the sparsity of the input frame in transform domain.\\

\item  The proposed framework gives better performance for the large resolutions for a same number of measurements compared to TV-DCT \cite{31}. However, a small  degradation of performance is observed for low resolution images. This may be attributed to absence of high-frequency information in the reconstructed images, particularly in low-resolution images. The reconstructed images  ($1/4^{th} $ of the input resolution) using BL+EL1 are the LL sub-band of the decomposition level 2, and the reconstructed images ($1/2^{th} $ of the input resolution) using BL+EL1+EL2 are the LL sub-band of the decomposition level 1. Reconstructed images (input resolution) using BL+EL1+EL2+EL3  provides the almost replica of the original image  and shows the higher performance.
\par In contrast to difficulty encountered in obtaining a suitable hardware implementation of the TV-DCT procedure \cite{31}, the proposed framework is amenable for hardware realization.

\item Finally, scalable features of the proposed framework are also verified for the possible three modes, viz. spatial, temporal and quality modes of scalability. Fig. \ref{result_image} shows the spatial scalability of the proposed framework. Three levels of decomposition have been performed for DWT, in which the third level LL sub-band coefficients are considered as a base layer, on the same level LH,HL and HH sub-bands of L-frame and all 4 sub-bands of $3^{rd}$ level in H-frame are considered as the enhancement layer 1 (EL1), second level as the enhancement layer 2 (EL2) and the first level constitute the enhancement layer 3 (EL3). Spatial scalability applied on three different full HD video sequences, resolution, PSNR and percentage of measurements used to reconstruct the frames for level 1 to level 3 (BL+EL1) are clearly given in Fig. \ref{result_image}. Frame rate of the base layer is $ 1: 2^{l} $, where $ l $ is the maximum number of levels used in the DWT. At most, three levels of temporal decomposition have been used in our architecture, so that the frame rate of the level $l$ is $ 1: 2^{l} $ of the original. The next higher resolution has the frame rate $ 1: 2^{l-1} $, and so on. In this way the proposed framework provides the temporal scalability.
\end{enumerate}

\section{Conclusions}

A framework for CS based VLSI friendly scalable video coding using 3-D DWT has been proposed in this paper. In this framework, 3-D DWT has been applied on the source video and measurements are made by using an adaptive measurement scheme (AMS). Wavelet transform at the encoder helps one calculate the exact sparsity in the wavelet domain, and AMS increases the compression ratio also reduces the complexity of the decoder. This paper also introduces the Enhanced Approximate Message Passing (EAMP) algorithm followed by the inverse DWT gives the reconstructed video sequence. The proposed EAMP provides the best results as compared to several other existing iterative algorithms with less number of measurements and iterations.   Formation of wavelet sub-bands through 3-D DWT and creating small input vectors of wavelet sub-bands through AMS, reduced the number of multipliers by 93$ \% $ in the proposed framework compared to conventional CS based video coding schemes, and made it more suitable for VLSI implementation. Scalable features of the DWT with different levels satisfies the requirements of the scalable video coding. The proposed framework is considered to be suitable for applications including space science, mobile streaming, wireless transmission and data compression by high speed cameras.\\


\vspace{-1.2cm}
\begin{IEEEbiography}[{\includegraphics[width=1in,height=1.1in]{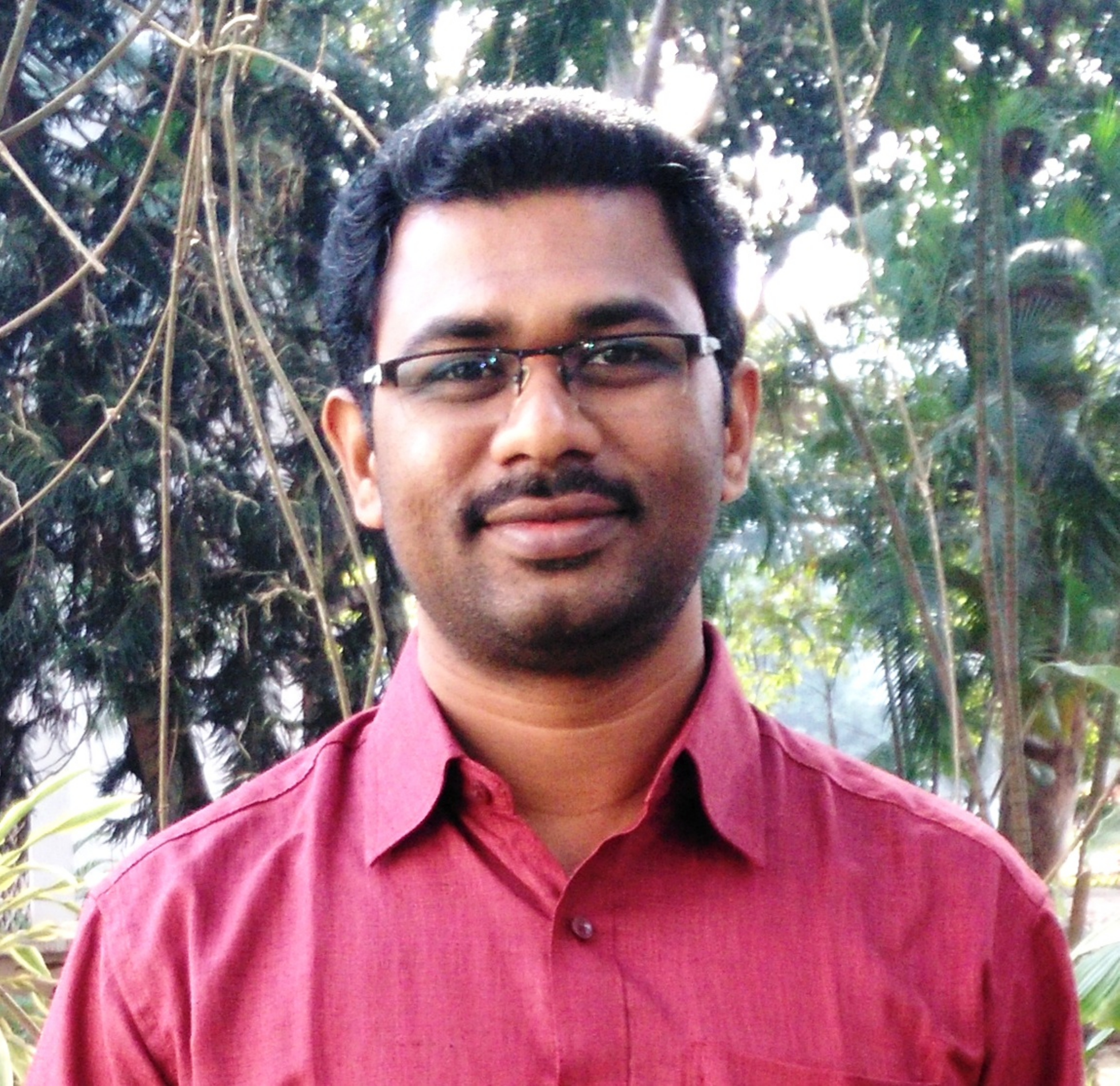}}]{Kota Naga Srinivasarao Batta}
  received the Master’s degree in Visual Information and Embedded Systems Engineering from IIT Kharagpur, India, in 2008. Currently he is working toward the Ph.D. degree in the department E$\&$ECE, IIT Kharagpur. From 2002 to 2006, he worked as an Assistant Professor and from 2008 to 2012 he worked as an Associate Professor in the department of ECE, Gudlavalleru Engineering College, Gudlavalleru, India. His research interests include VLSI architectures for image and video compression and design of embedded systems.
\end{IEEEbiography}
\vspace{-1.2cm}
\begin{IEEEbiography}[{\includegraphics[width=1in,height=1.1in]{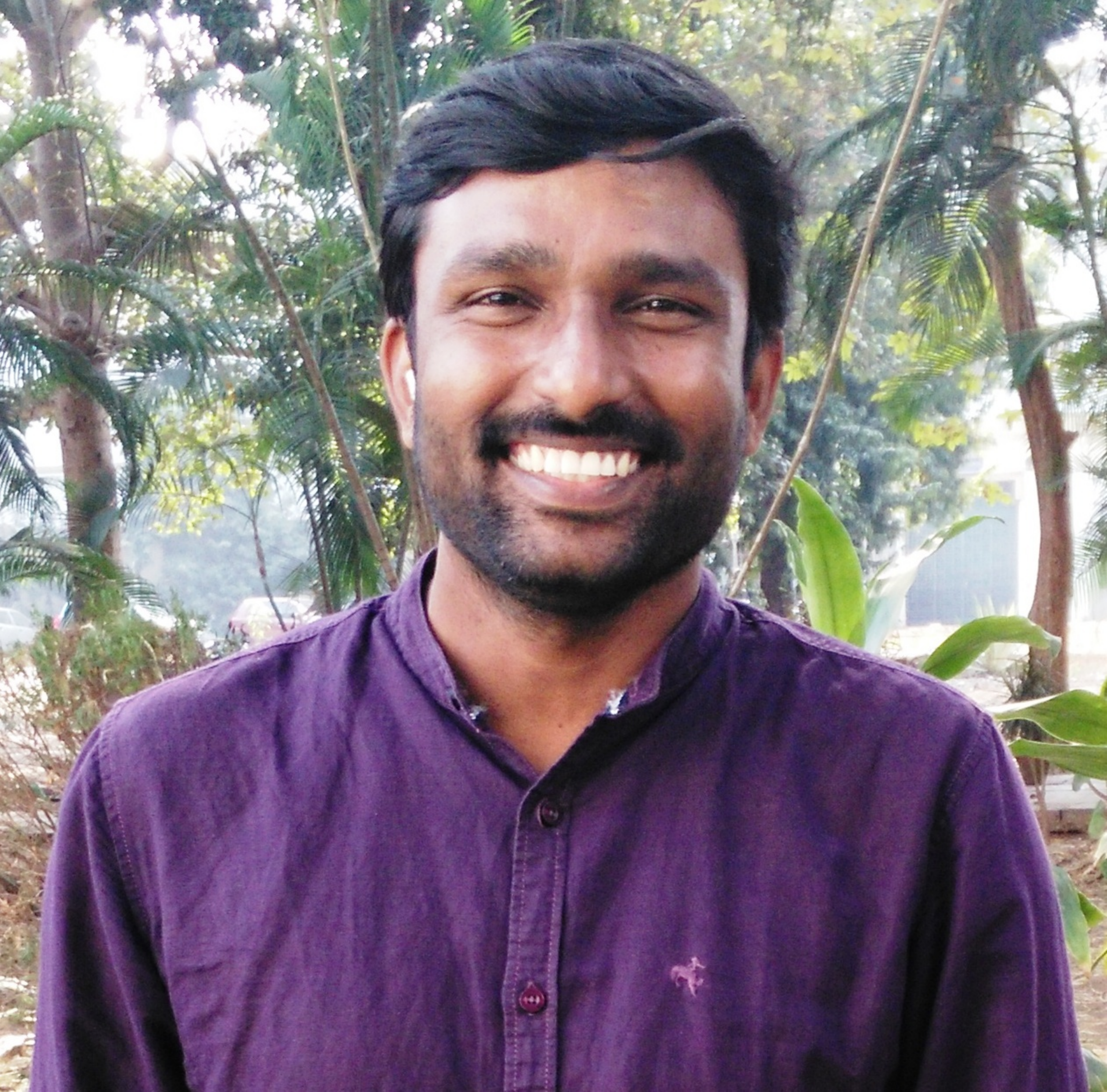}}]{Vinay Chakravarthi Gogineni}
    received the M.Tech.  degree in communications engineering  from VIT  Vellore, India, in 2008. From 2008 to 2011, he was with Cosyres technologies, Bangalore, where he was involved in algorithmic level optimization of audio codecs.  He is currently  working toward the PhD degree in the department of E$\&$ECE at IIT Kharagpur. His current research interests include adaptive filtering  and statistical signal processing.
\end{IEEEbiography}
\vspace{-1.2cm}
\begin{IEEEbiography}[{\includegraphics[width=1in,height=1.25in]{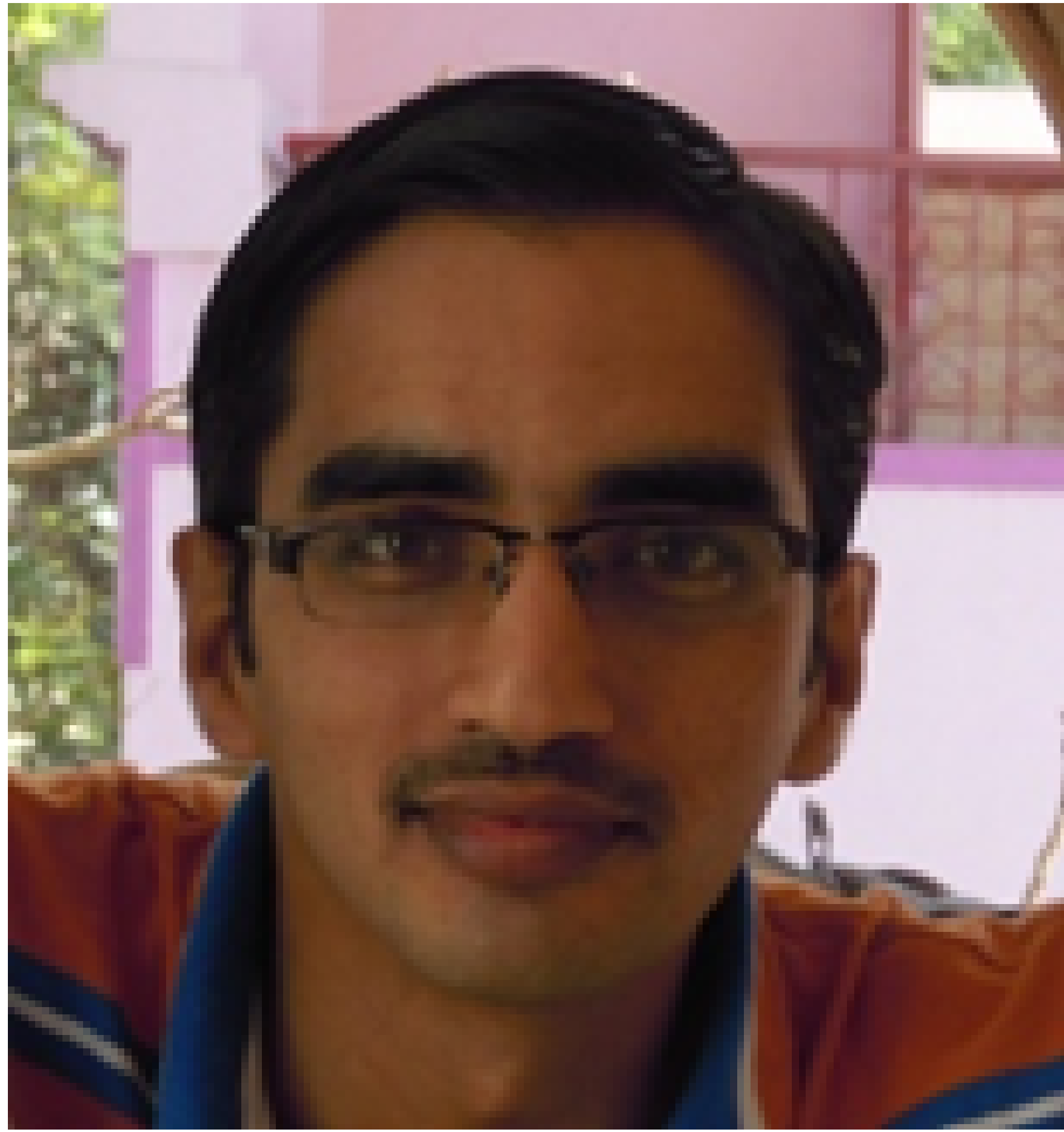}}]{Subrahmanyam Mula}
 received the M.Tech.  degree in microelectronics and VLSI design in 2003 from IIT Kharagpur, India, where he is currently working toward the Ph.D. degree in electronics engineering. From 2003 to 2014, he was with Intel  India, Bangalore, where he worked on architecture, implementation of test bench components and verification closure of complex modules , sub-systems and full-system of gigabit ethernet switches, processors  and GPUs.  His current research interests include VLSI signal processing and compressive sensing.
\end{IEEEbiography}
\vspace{-1.2cm}
\begin{IEEEbiography}[{\includegraphics[width=1in,height=1.25in]{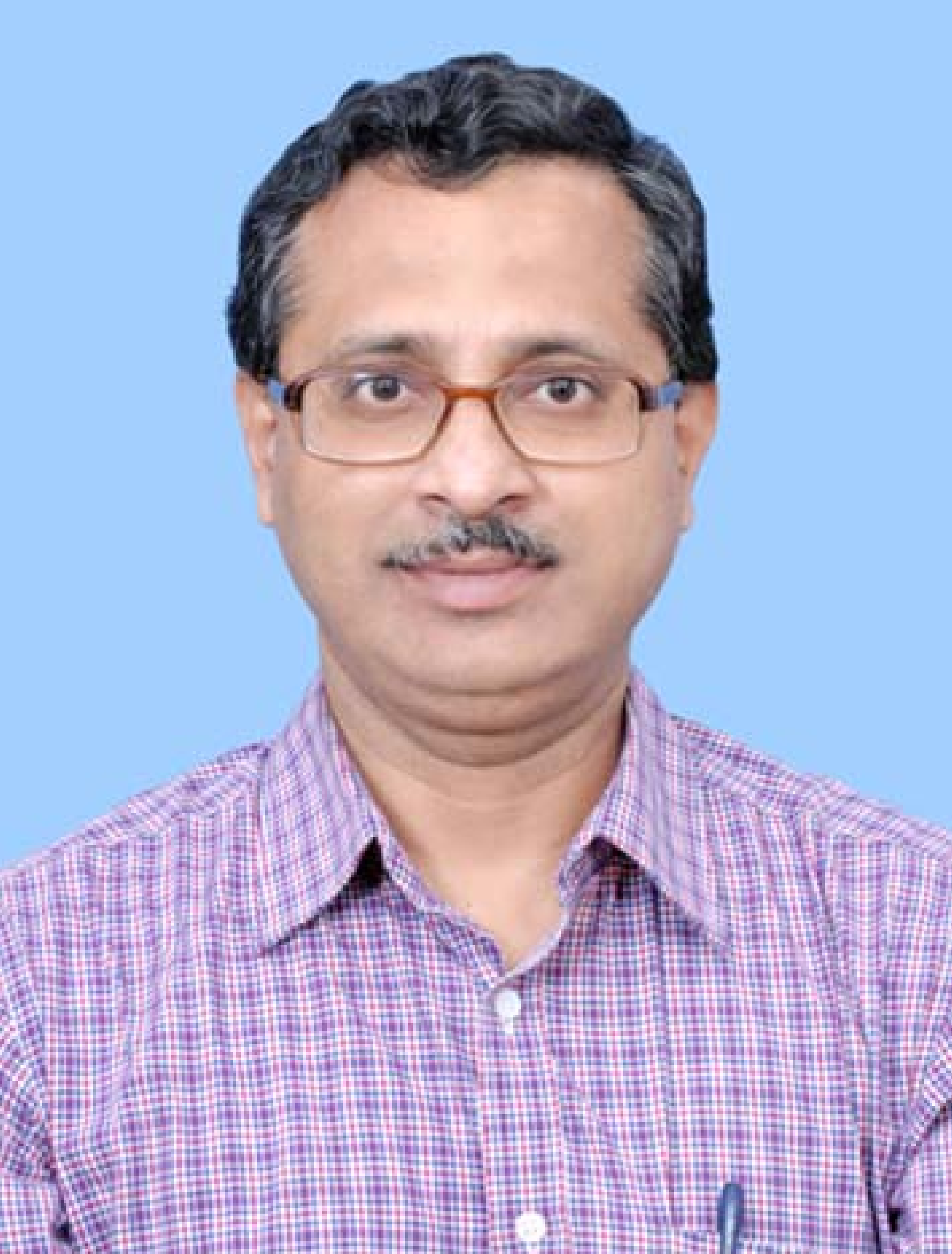}}]{Indrajit Chakrabarti}
  received the Ph.D from Indian Institute of Technology (IIT) Kharagpur, India in 1997. From 1998 to 2004, he worked as an Assistant Professor and later as an Associate Professor in the Department of Electronics and Communication Engineering, IIT Guwahati. He is presently serving as a Professor in the Department of Electronics and Communication Engineering, IIT Kharagpur. His research interests include VLSI architectures for image and video processing, digital signal processing, error control coding and wireless communication. He has published more than 20 papers in international journals, and is a member of IEEE.
\end{IEEEbiography}
\end{document}